\documentclass[prd,twocolumn,preprintnumbers,amsmath,amssymb,superscriptaddress,showpacs,showkeys,nofootinbib]{revtex4-1}
\usepackage{amsmath,amsfonts,amssymb,amsthm,amstext,amscd,eucal,srcltx,mathrsfs}
\usepackage{epsfig,graphicx,bm}
\usepackage{epstopdf, epsf}
\usepackage{dcolumn} 
\usepackage{hyperref}
\usepackage{braket}
\usepackage{color}
\newcommand{\be}{\begin{eqnarray}}
\newcommand{\ee}{\end{eqnarray}}
\renewcommand{\d}{\mbox{${\rm d}$}} 

\newcommand{\eg}{{\em e.g.}} 
\newcommand{\ie}{{\em i.e.}}

\newcommand{\lp}{\ell_{\rm P}}
\newcommand{\mpl}{m_{\rm P}}

\newcommand{\Rh}{R_{\rm H}}

\begin{document}
\title{The role of collapsed matter in the decay of black holes}
\author{Roberto Casadio}
\email{casadio@bo.infn.it}
\affiliation{Dipartimento di Fisica e Astronomia,
Alma Mater Universit\`a di Bologna,
via~Irnerio~46, 40126~Bologna, Italy}
\affiliation{I.N.F.N., Sezione di Bologna, IS FLAG
viale~B.~Pichat~6/2, I-40127 Bologna, Italy}
\author{Andrea Giusti}
\email{agiusti@bo.infn.it}
\affiliation{Department of Physics \& Astronomy,
Bishop's University, 
2600 College Street, J1M 1Z7 Sherbrooke, Qu\'{e}bec, Canada}
\begin{abstract}
We try to shed some light on the role of matter in the final stages of black hole evaporation
from the fundamental frameworks of classicalization and the black-to-white hole bouncing scenario. 
Despite being based on very different grounds, these two approaches 
attempt at going beyond the background field method and treat black holes as fully quantum
systems rather than considering quantum field theory on the corresponding classical manifolds.
They also lead to the common prediction that the semiclassical description of black hole evaporation
should break down and the system be disrupted by internal quantum pressure,
but they both arrive at this conclusion neglecting the matter that formed the black hole.
We instead estimate this pressure from the bootstrapped description of black holes, which allows us
to express the total Arnowitt-Deser-Misner mass in terms of the baryonic mass still present
inside the black hole.
We conclude that, although these two scenarios provide qualitatively similar predictions for the final stages,
the corpuscular model does not seem to suggest any sizeable deviation from the semiclassical
time scale at which the disruption should occur, unlike the black-to-white hole bouncing scenario.
This, in turn, makes the phenomenology of corpuscular black holes more subtle from an astrophysical
perspective.
\end{abstract}
\maketitle
\section{Introduction}
The possible existence of black holes is one of the most striking predictions of General Relativity
and understanding their true nature has become one of the hottest topics of current theoretical research
after the detection of gravitational waves and the reconstruction of their ``shadow''.
In particular, the famous result that black holes could evaporate~\cite{hawking} raised a number of
paradoxes and made it apparent that a consistent quantum description of strong gravitational fields might
require going beyond the (up to then) very successful application of quantum field theory on classical curved
backgrounds.
A different approach is certainly necessary in order to understand the late stages of the evaporation
(see, {\em e.g.}~\cite{bardeen} and references therein).
\par
A quantum bounce is a typical scenario which emerges when one tries to include quantum mechanical
effects in the description of self-gravitating systems (see \eg~\cite{Casadio:1998yr,Kamenshchik:2018rpw}
and references therein).
Thus, it is not hard to believe that this sort of effects should turn out to be massively relevant in the very late
stages of the life of a black hole.
On similar considerations are based the models of the Planck star~\cite{Rovelli:2014cta} and of
the black-to-white hole quantum transitions~\cite{Haggard:2014rza}.
More specifically, the fundamental idea is that classical General Relativity becomes unreliable 
when the system enters regimes of very high curvature, such as regions close to black hole singularities.
In a Planck star, the classical metric manifold inside regions of large curvature is therefore replaced by
a fully quantum space-time as it would emerge in Loop Quantum Gravity.
One naively expects that quantum mechanical effects will dominate at a scale $\ell \sim \lp$, where $\lp$
denotes the Planck length.
However, if one considers the typical bouncing scenarios within the mini-superspace approach
to quantum cosmology~\cite{Novello:2008ra}, the general conclusion is that the bounce
should occur when a certain model-dependent {\em critical density} $\rho _{\rm c}$
is reached.
By applying this result to the gravitational collapse, one expects that the core of the
system, where the matter is supposed to end after crossing the event horizon, stops contracting
at a comparable critical density $\rho _{\rm c}$.
For example, in Loop Quantum Cosmology~\cite{Agullo:2016tjh}, one finds
$\rho _{\rm c} \simeq \rho _{\rm P} \simeq \mpl / \lp ^3$, where $\mpl$ is the Planck mass.
Hence, in spherical symmetry, the size of the inner quantum mechanical region,
representing the core of a Planck star, is naively given by $\ell \sim (M / \mpl) ^{1/3} \lp$, with $M$ denoting
the Arnowitt--Deser--Misner (ADM) mass~\cite{adm} of the system.
Notice that $\ell$ is several orders of magnitude larger than $\lp$ for an astrophysical black hole,
which, therefore, offers a window for detecting quantum gravitational effects at scales far from the
Planck regime.
If we buy this argument, the semiclassical picture of black hole evolution driven
by the Hawking effect should proceed smoothly up to the scale $\ell$, at which point the horizon
hits the boundary of the quantum gravity region and the quantum pressure tears the core apart,
thus freeing up what remains of its energy content.
\par
The natural extension of this very simple argument consists in describing the final stage of
black hole evaporation as a tunneling process from a black hole geometry to a white hole
geometry~\cite{Haggard:2014rza}.
More precisely, in this model the two classical space-time patches are glued together through
an intermediate region, dominated by quantum gravitational effects, resembling a sort of
Oppenheimer-Snyder collapse~\cite{Oppenheimer:1939ue} with an extra bounce into a
white hole~\cite{Haggard:2014rza, Malafarina:2018pmv} 
(see also~\cite{hajicek} and references therein for similar results in canonical Einstein gravity).
One of the most interesting results following from this effective description of the tunneling
process is that the quantum bounce might actually be realized after a time 
\be
\tau_{\rm b}
\sim
\lp \, \frac{M^2}{\mpl ^2}
\ ,
\label{tb}
\ee
which is way shorter than the standard Hawking evaporation time 
\be
\tau _{\rm H}
\sim
\lp \,
\frac{M^3}{\mpl ^3}
\ .
\label{th}
\ee
In the semiclassical picture, black hole should evaporate semiclassically up to the Page time~\cite{bardeen}.
If we trust the Hawking formula for a time $\tau \sim \tau_{\rm b}$, its total mass at the time of the bounce is given by 
\be
M_{\rm b}\simeq M \, \sqrt[3]{1- (\mpl / M)}
\simeq
M
\ ,
\ee
which means that the entire black hole mass ``bounces out'' for $M \gg \mpl$.
\par
Several phenomenological studies concerning these loopy bouncing black holes have been carried
out in the last few years (see \eg~\cite{Barrau:2015uca, Vidotto:2018wvr} and references therein),
with a natural focus on the final stages of evolution of primordial black holes.
In these studies it was found that the mass of primordial black holes ``exploding'' today,
after a time $\tau \sim \tau_{\rm b}$ from formation, should range from $10^{11} \, {\rm kg}$
to $10^{20} \, {\rm kg}$~\cite{Rovelli:2014cta, Barrau:2015uca}.
The aim of this letter is to discuss the bouncing scenario of black--to--white hole transitions
within the framework of the classicalization scheme~\cite{Gia0,Gia1} and of the corpuscular
picture of gravity~\cite{baryons, planckian} by taking explicitly into account the role of matter
collapsed inside the black hole~\cite{toappear}. 
\section{Corpuscular gravity}
The goal of classicalization~\cite{Gia0} is to tackle the problem of the UV completion of effective field theories
without forcing in any new (hard) degree of freedom, as it would happen in the Wilsonian approach.
This idea lies at the very foundation of the corpuscular model of black holes,
also known as black hole's quantum $N$-portrait~\cite{Gia1}.
According to this view, a black hole can be described as a leaky bound state of a large number
$N_{\rm G}$ of soft virtual gravitons.
In particular, while the collective gravitational coupling
\be
g
=
N_{\rm G} \, \alpha \simeq 1
\label{g}
\ee
and the system is {\em globally\/} in the strong coupling regime, the effective coupling among the constituents
\be
\alpha
\simeq
\frac{1}{N_{\rm G}}
\label{alpha}
\ee
remains extremely small, which suppresses the contribution of loop corrections {\em locally}.
It is also worth recalling, for the sake of argument, that in this framework one finds the fundamental scaling
relations
\be
\label{scaling-dvali}  
M
\simeq
\sqrt{N_{\rm G}} \, \mpl
\ , 
\qquad
\varepsilon
\simeq
\frac{\mpl}{\sqrt{N_{\rm G}}}
\ ,
\ee
where $\varepsilon$ denotes the typical energy of each graviton in the bound state,
provided the typical length scale of all the constituents is $\lambda \simeq \Rh$
(see, \eg~\cite{review} and references therein).
It is then easy to conclude that the classical description of gravity can emerge
naturally from this picture as a direct consequence of the large value of the occupation
number $N_{\rm G}$, ultimately leading to an effective classical behaviour.
\par
On the other hand, semiclassical effects like the Hawking process and the generalized second law
of black hole thermodynamics arise inherently from the softness of the constituents and their
combinatorics.
Indeed, in this scenario the Hawking radiation is understood as the leakage of the bound state
due to $2 \to 2$ scattering processes among the gravitons in the system.
The total depletion rate caused by the scattering of each graviton with the other $N_{\rm G}-1$
constituents is roughly given by~\cite{Gia1, baryons}
\be 
\label{depletion}
\notag
\Gamma 
&\simeq& 
\alpha ^2\, N_{\rm G} \, (N_{\rm G}-1) \, \frac{\varepsilon}{\hbar}
+ \mathcal{O} \left( \frac{1}{\lp \, N_{\rm G}^{3/2}} \right)
\\ 
&\simeq& 
\frac{1}{\sqrt{N_{\rm G}} \, \lp}
+ \mathcal{O} \left( \frac{1}{\lp \, N_{\rm G}^{3/2}} \right)
\ ,
\ee
where $\varepsilon$ is, indeed, the typical energy involved in this process.
On recalling the scaling laws~\eqref{scaling-dvali} and that $\dot N_{\rm G} \simeq - \Gamma$,
this yields
\be
\varepsilon\,\dot N_{\rm G}
\sim
\dot M
\sim
-\frac{\mpl}{\lp}
\left(\frac{\mpl}{M}\right)^2
\ ,
\label{dotM}
\ee
and we conclude that, in the corpuscular picture, the (apparent) thermal behaviour of the Hawking radiation
follows from the softness of the gravitons leaving the system and from their combinatorics.
It is then important to recall that one also expects a flux of comparable energy for ordinary matter from the
scattering of baryons by gravitons~\cite{baryons}.
Consequently, the description of Hawking evaporation as quantum depletion allows for a natural resolution
of the information loss problem since the features of the collapsed matter are stored within the system,
for a very long time, before they start to gradually leak out.
In other words, the unitarity of the time evolution should be restored through the emission of quantum
hair~\cite{quantumhair},
which is subjected to a suppression of order $1/M^2$, contrary to the standard semiclassical picture in which
these fields are exponentially faint.
\par
From this quantum field theoretic description of black holes and the Hawking radiation we also get a taste
of when such a model should brake down.
A careful inspection of Eq.~\eqref{depletion} tells us that the quantum gravitational corrections
to the semiclassical picture of gravity should become relevant when the condition $N_{\rm G} \gg 1$
ceases to be valid.
In other words, when the number of gravitons in the black hole has become small enough that the collective
gravitational interaction~\eqref{g} can no more overcome the increasing quantum pressure,
the system will not remain confined within the Schwarzschild radius 
\be
\Rh 
=
2\,\lp\,\frac{M}{\mpl}
\simeq
\sqrt{N_{\rm G}} \, \lp
\ .
\ee
The bound state should break off at that point, freeing up all the remaining matter and gravitons.
For a ``purely gravitational'' black hole made only of gravitons, we expect that this occurs for
$N_{\rm G}$ of order one, when $\Rh\sim \lp$.
However, if one realistically requires the presence of regular baryonic matter up to the latest
stages of the evaporation, the typical values of $N_{\rm G}$ at which the semiclassical behaviour breaks
down could be much bigger. 
\par
We have seen that corpuscular black holes share some similarities with the typical dynamics expected
for Planck stars.
One can then investigate the phenomenological implications for primordial black holes, considering the scenario
of black-to-white hole transitions, if we add the classicalization scheme to the mix.
For example, the smallest possible mass of a primordial black hole exploding today should be
$M \simeq 10^{11} \, {\rm kg}$ according to Ref.~\cite{Vidotto:2018wvr}.
From Eq.~\eqref{scaling-dvali} one immediately finds $N_{\rm G} \simeq 10^{38} \gg 1$ at the time of the
bounce.
According to the black-to-white hole picture, the semiclassical behaviour should therefore break down 
at scales $\ell\sim 10^{19}\,\lp$, about a tenth of the proton's size.
\par
The huge discrepancy between $N_{\rm G}\sim 1$ and $N_{\rm G}\sim 10^{38}$ at the time of the departure from 
the semiclassical evolution estimated in the two scenarios leads us to the fundamental questions:
What is the shortest scale of gravitational confinement for ordinary matter?
In order to answer this question, we need an estimate of the quantum pressure that prevents the baryons
from reaching the infinitely dense classical singularity~\cite{boot,brustein}.
\section{Bootstrapped black holes}
The bootstrapped description~\cite{boot} of stars and black holes~\cite{toappear} was recently suggested as
an effective realisation of the classicalization scheme for the gravitational interaction.
This approach is simply constructed by introducing the leading order non-linearities predicted by
General Relativity in Newtonian gravity. 
The effective Lagrangian for the gravitational potential is found to be~\cite{toappear}
\be
L[V]
&=&
L_{\rm N}[V]
-4\,\pi
\int_0^\infty
r^2\,\d r
\left[
J_V\,V
+
J_\rho\left(\rho+p\right)
\right]
\nonumber
\\
&=&
-4\,\pi
\int_0^\infty
r^2\,\d r
\left[
\frac{\mpl\left(V'\right)^2}{8\,\pi\,\lp}
\left(1-4\,V\right)
\right.
\nonumber
\\
&&
\left.
\phantom{-4\,\pi\int_0^\infty r^2\,\d r\ }
+\left(\rho+p\right) V\left(1-2\,V\right)
\right]
\ ,
\label{LagrV}
\ee
where 
\be
L_{\rm N}[V]
=
-4\,\pi
\int_0^\infty
r^2 \,\d r
\left[
\frac{\mpl\left(V'\right)^2}{8\,\pi\,\lp}
+\rho\,V
\right]
\label{LagrNewt}
\ee
is the Newtonian part, 
\be
J_V
=
-\frac{\mpl\left( V' \right)^2}{2\,\pi\,\lp}
\label{JV}
\ee
is the gravitational current and $J_\rho=-2\,V^2$ the higher order correction to the matter part.
The corresponding field equation
\be
\triangle V
=
4\,\pi\,\frac{\lp}{\mpl}\left(\rho+p\right)
+
\frac{2\left(V'\right)^2}
{1-4\,V}
\label{EOMV}
\ee
and the conservation equation 
\be
p'
=
-V'\left(\rho+p\right)
\ ,
\label{eqP}
\ee
allow for finding explicit (classical) solutions generated by arbitrarily compact matter sources. 
In fact, it was recently shown that the baryonic pressure can, in principle, counterbalance the gravitational 
pull for any radius $R$ of the matter source in this framework~\cite{toappear}.
In other words, there is no Buchdahl limit and, for $R\lesssim \Rh$, one also finds that the total proper mass
of $N_{\rm B}$ baryons $M_0=N_{\rm B}\,\mu$ must relate with the ADM mass
according to~\footnote{For a more accurate estimate of the scaling for $R\simeq \Rh$ see Ref.~\cite{toappear}.
Given the models considered here are still rather unsophisticated and (more importantly) the qualitative
nature of the fundamental questions we address, all calculations will remain at the level of order
of magnitude estimates}.
\be
\frac{\lp\,M_0}{R\,\mpl}
\sim
\left(\frac{\lp\,M}{R\,\mpl}\right)^\alpha
\ ,
\label{M0M}
\ee
where $\alpha=2/3$ for a homogeneous matter distribution.
For the sake of generality, we will just assume $0<\alpha<1$.
First of all, we need
\be
R\lesssim \Rh
\simeq
2\,\lp\,\frac{M}{\mpl}
\ ,
\ee
in order for the system to be a black hole.
For the initial configuration, we can assume $R\ll \Rh$ and the baryons are in a highly relativistic regime,
so that the initial size of the baryon source
\be
R_{\rm in}
\sim
\lambda_{\rm B}
\sim
\lp\,\frac{N_{\rm B}\,\mpl}{M_{\rm in}-M_0}
\sim
\lp\,\frac{N_{\rm B}\,\mpl}{M_{\rm in}}
\ ,
\label{Rin}
\ee
in which we used $M_{\rm in}\gg M_0$ for $R\ll \Rh$, as follows from Eq.~\eqref{M0M}.
Due to the Hawking evaporation of gravitons, the ADM mass $M$ decreases.
Since the (initial) Hawking temperature is very low for astrophysical black holes, 
we can safely assume no baryon is emitted and $M_0$ remains constant.
From Eq.~\eqref{M0M} we then infer that
\be
\frac{\Delta R}{R}
\simeq
-\frac{\alpha}{1-\alpha}
\frac{\Delta M}{M}
>0
\ ,
\ee
and the size of the baryon source inside the black hole increases while the hole
evaporates semiclassically.
\par
Let us then consider a final configuration in which the size of the source
$R_{\rm fin}\sim \Rh$, or, again from Eq.~\eqref{M0M}, 
\be
M_{\rm fin}\simeq M_0
\ .
\label{Mfin}
\ee
In this case, the baryons are no more highly relativistic and 
\be
R_{\rm fin}
\sim
\lambda_{\rm B}
\sim
\lp\,\frac{\mpl}{\mu}
\ ,
\label{Rfin}
\ee
where $\mu$ is the proper mass of one of the $N_{\rm B}$ baryons.
Clearly, after this point, the size of the baryon source could exceed the
gravitational radius and the object would not be a black hole any more.
From Eqs.~\eqref{M0M} and \eqref{Rin}, we also have
\be
\notag
\frac{M_0}{\mpl}
&\sim&
\left(\frac{R_{\rm in}}{\lp}\right)^{1-\alpha}
\left(\frac{M_{\rm in}}{\mpl}\right)^{\alpha}
\\
&\sim&
N_{\rm B}^{1-\alpha}
\left(\frac{M_{\rm in}}{\mpl}\right)^{2 \alpha - 1}
\ ,
\ee 
or
\be
\frac{M_{\rm in}}{\mpl}
&\sim&
\left(
\frac{M_0}{\mu}
\right)^{\frac{\alpha}{2\,\alpha-1}}
\left(
\frac{\mu}{\mpl}
\right)^{\frac{1}{2 \alpha - 1}}
\nonumber
\\
&\sim&
N_{\rm B}^{\frac{\alpha}{2\,\alpha-1}}
\left(
\frac{\mu}{\mpl}
\right)^{\frac{1}{2 \alpha - 1}}
\ ,
\label{MinMfin}
\ee
which depends on the baryon mass $\mu$ and number of baryons $N_{\rm B}$.
\par
The evaporation process is governed by the master equation~\eqref{dotM},
which leads to an evaporation time
\be 
\tau _{\rm {in} \to {fin}}
\sim
\lp 
\left(\frac{M_{\rm in}}{\mpl}\right)^3
\left[1-
\left(\frac{M_{\rm fin}}{M_{\rm in}}\right)^3
\right]
\ .
\ee
In the Hawking picture one usually considers complete evaporation, {\em i.e.}, $M_{\rm fin} = 0$,
thus leading to the well-known result~\eqref{th}, or more precisely
\be
\label{time-H}
\tau _{\rm H}
\sim
\lp
\left( \frac{M_{\rm in}}{\mpl} \right)^3
\ .
\ee 
On the other had, we just saw that $M_{\rm fin}\sim M_0$ in the bootstrapped picture
and Eq.~\eqref{MinMfin} yields
\be
\frac{M_{\rm fin}}{M_{\rm in}}
&\sim&
\left(
\frac{\mpl^2}{\mu\,M_{\rm fin}}
\right)^{\frac{1-\alpha}{2\,\alpha-1}}
\nonumber
\\
&\sim&
\left(
\frac{\mpl^2}{\mu\,M_{\rm in}}
\right)^{\frac{1-\alpha}{\alpha}}
\ .
\ee
The above ratio must be smaller than one,
which requires $N_{\rm B}>\mpl^2/\mu^2$
(equivalent to $M_0/\mpl>\mpl/\mu$),
if $1/2<\alpha<1$ or $N_{\rm B}<\mpl^2/\mu^2$
(equivalent to $M_0/\mpl<\mpl/\mu$)
if $0<\alpha<1/2$.
Under these conditions, we then obtain the evaporation time
\be
\label{time-a}
\tau _{\alpha} 
&\sim& 
\tau_{\rm H}
\left[
1
-
\left( \frac{\mpl ^2}{\mu\,M_{\rm fin}} \right)^{\frac{3\,(1-\alpha)}{2\,\alpha-1}}
\right]
\nonumber
\\
&\sim& 
\tau_{\rm H}
\left[
1
-
\left( \frac{\mpl ^2}{\mu\,M_{\rm in}} \right)^{\frac{3\,(1-\alpha)}{\alpha}}
\right]
\ .
\ee
This expression tells us that $\tau _{\alpha} < \tau_{\rm H}$, but the two
times differ significantly only provided $\mu\,M_{\rm fin}\sim \mpl^2$.
For instance, if we require $\tau _{\alpha}$ equals the Page time
$\tau_{\rm Page}\simeq (7/8)\,\tau_{\rm H}$ at which
$M_0\sim M_{\rm fin}\simeq M_{\rm in}/2$, 
we obtain
\be
\left( \frac{\mpl ^2}{\mu\,M_{\rm fin}} \right)^{\frac{1-\alpha}{2\,\alpha-1}}
\simeq
\frac{1}{2}
\simeq
\left(
\frac{\mpl^2}{\mu\,M_{\rm in}}
\right)^{\frac{1-\alpha}{\alpha}}
\ .
\ee
For a source made of neutrons with mass $\mu\simeq 10^{-19}\,\mpl$,
for $1/2<\alpha<1$, we find
\be
M_{\rm fin}
\simeq
2^{\frac{2\,\alpha-1}{1-\alpha}}\cdot 10^{19}\,\mpl
\simeq
2^{\frac{2\,\alpha-1}{1-\alpha}}\cdot
10^{-19}
\,
M_\odot
\ ,
\ee
and
\be
M_{\rm in}
\simeq
2^{\frac{\alpha}{1-\alpha}}\cdot
10^{19}\,\mpl
\simeq
2^{\frac{\alpha}{1-\alpha}}\cdot 
10^{-19}
\,
M_\odot
\ ,
\ee
where $M_\odot \sim 10^{38} \, \mpl$ is the solar mass.
\par
For the particular case $\alpha=2/3$, the above expressions
become
\be
M_{\rm fin}
\simeq
2\cdot 10^{19}\,\mpl
\simeq
2\cdot 
10^{-19}
\,
M_\odot
\simeq
0.5\,M_{\rm in}
\ ,
\ee
which corresponds to an object of final radius
\be
R_{\rm fin}\sim\lambda_{\rm B}\simeq 10^{19}\,\lp\simeq 10^{-16}\, {\rm m}
\ .
\ee
This result is in line with the idea that the breakdown of the semiclassical evaporation
should happen at a scale way above the Planck regime, as suggested by the loop-inspired
scenario.
However, if one tries to put together the corpuscular picture of gravity with the role of matter
in the late stages of the evaporation, the expected disruption of the system is found to occur around
the Page time~\cite{bardeen}, that is for $\tau_{\rm Page}\sim M^3-(M/2)^3 \sim M^3\sim \tau _{\rm dep}$.
Yet, this result is more in agreement with the semiclassical scenario.
\section{Concluding remarks}
In this work, we have tried to put the role of matter in the final stages of a black hole life under the spotlight. 
To do that, we have compared two quantum models for black holes, namely the corpuscular theory
and the black-to-white hole transition. 
The first thing that we have been able to observe is that, despite emerging from two completely different 
background pictures, they seem to predict a common scenario, namely that the semiclassical picture will
necessarily brake down and the system will be disrupted by internal quantum pressure. 
The key difference between these two pictures resides in the timescale at which this effect should occur. 
Indeed, at least from the perspective of Loop Quantum Gravity, the black-to-white hole transition is expected 
after a time $\tau \sim M^2$ from the black hole formation.
In the corpuscular picture instead the depletion time scales like the Hawking time, \ie,~$\tau \sim M^3$,
since there should be no physical reasons, at least in this effective field theory of gravity,
for the emergent semiclassical picture to fail before that.
\par
This conclusion makes the phenomenology of the corpuscular scenario more subtle to test
from an astrophysical perspective because the typical time scales are close to those predicted
by the semiclassical description.
Clearly, some different signature is required in order to reveal the quantum nature of black holes
and the theoretical investigation of all possible models needs to be continued.
\subsection*{Acknowledgments}
We would like to thank G.~Dvali and A.~Giugno for useful comments.
This work was partially supported by the INFN grant FLAG, and it has also been carried out in the framework
of the activities of the National Group for Mathematical Physics (GNFM, INdAM).
\end{document}